\documentclass[11pt,twoside]{atmp}

\usepackage{amsmath,amssymb}

\usepackage[all]{xy}
\usepackage{color}
\usepackage{graphicx} 

\begin{document}

\title{The edge spectrum of Chern insulators with rough boundaries}

\author{Emil Prodan}

\address{Department of Physics, Yeshiva University, New York, NY 10016}

\date{\today}
\maketitle

\begin{abstract}

Chern insulators are periodic band insulators with the property that their projector onto the occupied bands have non-zero Chern number. Chern insulator with a homogeneous boundary display continuum spectrum that fills the entire insulating gap. The local density of states corresponding to this part of the spectrum is localized near the boundary, hence the terminology edge spectrum. An interesting question arises, namely, if a rough boundary, which can be seen as a strong random potential acting on these quasi 1-dimensional states, would destroy the continuum edge spectrum. This paper shows how such question can be answered via a newly formulated abstract framework in which the expectation value of the current of a general observable is connected to the index of a specific Fredholm operator. For the present application, we will connect the expectation value of the charge edge current with the index of a Fredholm operator that remains invariant under arbitrary deformations of the boundary. 
\end{abstract}

\section{Introduction} 

For magnetic Schrodinger operators in half-plane, Hatsugai established a fundamental result \cite{Hatsugai:1993cs,Hatsugai:1993jt} that says that the number of conducting channels forming in the gap of the bulk system due to the presence of an edge is equal to total Chern number of the bands below that gap. The work by Hatsugai applies only for rational magnetic fluxes and  homogeneous edges with Dirichlet boundary conditions. Almost 10 years later, using an advance mathematical machinery, Kellendonk, Richter and Schulz-Baldes established \cite{Kellendonk:2002p598} a new link between the bulk and edge theory, which ultimately allowed them to generalize Hatsugai's statement to half-plane magnetic Schrodinger operators with weak random potentials, irrational magnetic fluxes and general boundary conditions (the result was first announced in Ref.~\cite{SchulzBaldes:2000p599}). The result was later extended to continuous magnetic Schrodinger operators in Ref.~\cite{Kellendonk:2004p597}. These breakthrough results are especially important since they directly relate the quantization of the edge currents to a new topological invariant, the index of a specific Fredholm operator. Using this new invariant, one can directly explore the topology of the edge states under quite general (and physically relevant) boundary conditions, without the need of any gedanken experiments that employ artificial boundary conditions.  

Under similar assumptions, the equality between bulk and edge Hall conductance was also demonstrated by  Elbau and Graf, soon after the publication of Ref.~\cite{Elbau:2002qf}, this time using more traditional methods.  In Ref.~\cite{Combes:2005qd}, which treats continuous magnetic Schrodinger operators, the edge appears at the separation between a left and a right potential, potentials that can assume quite general forms, in particular, they can include strong disorder. A similar result was established for discrete Schrodinger operators in Ref.~\cite{Elgart:2005rc}. We should mentioned that certain regularization of the edge conductance was necessary for the case of strong disorder. It is also important to notice that these papers \cite{Elbau:2002qf,Combes:2005qd,Elgart:2005rc} give a direct relation between the edge and bulk conductance and they do not discuss and edge index. However, in many situations, the edge index introduced in Ref.~\cite{Kellendonk:2004p597} is quite useful \cite{Prodan:2008oq} and this is precisely what we will discuss in this paper.

The works we mentioned above studied the Integer Hall effect for simple magnetic Shcrodinger operators in half-plane. In 1988, Haldane introduced a new model that exhibits the Integer Quantum Hall effect even thought the net magnetic flux per primitive cell is zero \cite{Haldane:1988dz}. This strange property is typically present in tight-binding models with more than one quantum state per site and with broken time-reversal symmetry. It is a consequence of the non-trivial Chern number associated to the occupied bands of the model. This work lead to the discovery of a new class of insulators that now bear the name Chern insulators. It was argued in the literature \cite{Sheng:2006vn,Thonhauser:2006kx} that,  due to their non-trivial topological properties,  finite samples of Chern insulators must posses edge conducting channels that wind around the sample similar to what happens in the magnetic case. Thus, Chern insulators will display remarkable edge physics, without the need of a macroscopic magnetic field. We should mention that no crystals of Chern insulators were discovered so far. It was argued, however, that by patterning ferroelectric materials one could fabricate a photonic crystal \cite{Haldane:2008ys} with similar properties.

Theoretically, Chern insulators are as ubiquitous as the normal insulator. Generally, thight-binding Hamiltonians with more than one quantum state per site and with broken time reversal symmetry can be always tuned to exhibit bands with non-zero Chern number. One of the simplest example of a Chern insulator is given by
\begin{equation}
\begin{array}{c}
H=\sum\limits_{{\bf n},i=0,1,2}\{|{\bf n},1\rangle \langle {\bf n}+{\bf b}_i,2|+|{\bf n}+{\bf b}_i,2\rangle \langle {\bf n},1| \} \medskip \\
+\sum\limits_{{\bf n},i=1,2}\{t |{\bf n},1\rangle \langle {\bf n}+{\bf b}_i,1|+t^*|{\bf n}+{\bf b}_i,1\rangle \langle {\bf n},1|  \medskip \\
+t^* |{\bf n},2\rangle \langle {\bf n}+{\bf b}_i,2|+t|{\bf n}+{\bf b}_i,2\rangle \langle {\bf n},2| \},
\end{array}
\end{equation}
where ${\bf n}$ labels the sites of a planar lattice generated by ${\bf b}_1$ and ${\bf b}_2$, ${\bf b}_0$=0 and Im$[t]$$\neq$0. This model is unitarily equivalent to the original Haldane's model and, in the bulk, it exhibits two bands with Chern numbers equal to $\pm 1$. The edge structure of the model was studied in great detail in Ref.~\cite{Thonhauser:2006kx} and the numerical calculations show one edge channel forming along the edge of the sample, in accordance to what one will predict from the topology of the bulk bands (in these calculations one rather sees two channels since the calculations were performed for slabs rather than half-plane samples). The calculations were, of course, performed for homogeneous edges. Our main question here is what happens to the edge channels for a  rough boundary. The question is interesting since the edge channels are localized near the boundary and the rough boundary can be seen as a strong random potential acting on these quasi-onedimensional states.

While the question is interesting in itself, by solving it we want to demonstrate an application of the abstract machinery given in Ref.~\cite{Prodan:2008oq}, where we developed an axiomatic framework for current quantization. In the process, we identified the minimal conditions in which such quantization takes place. We claim that this minimal conditions bring significant simplifications over the previous works and the main goal of the present paper is to show how standard it is to demonstrate that these minimal conditions are met in Chern insulators. For example, in  Ref.~\cite{Kellendonk:2004p597}, the authors rely heavily on the Gaussian decay of the heat kernel for bulk magnetic Schrodinger operators and the estimates throughout the paper are quite involved. Here we will show that just simple exponential decay estimates on the bulk resolvent will do the job. Moreover, we will demonstrate that, once these exponential decay estimates are established, the minimal conditions stated in Ref.~\cite{Prodan:2008oq} can be derived by following several standard, model independent steps. Thus, the present application completes the abstract formalism of Ref.~\cite{Prodan:2008oq} by showing that the required minimal conditions can be demonstrated via a somewhat universal technique. We hope that the present simplifications will make the technique more accessible to the theoretical physicists and that the generality of the proofs will encourages a search for new systems where the technique could lead to non-trivial insight.

\section{Definition of the physical systems}

In this section we define a bulk system, whose dynamics is determined by a very general tight-binding Hamiltonian $H_0$. Then we define and parametrize what we call a rough boundary $\Gamma$.  A probability measure is introduced over the set of all admissible boundaries, which will be used for averaging. An auxiliary Hamiltonian $H$ and the Hamiltonian $H_\Gamma$ for the system with a boundary $\Gamma$ is also introduced in this Section. We describe the most important properties of these systems, whose complete derivations will be given in the Appendix. 

\subsection{The bulk system.}

\begin{figure}
 \center
 \includegraphics[width=7cm]{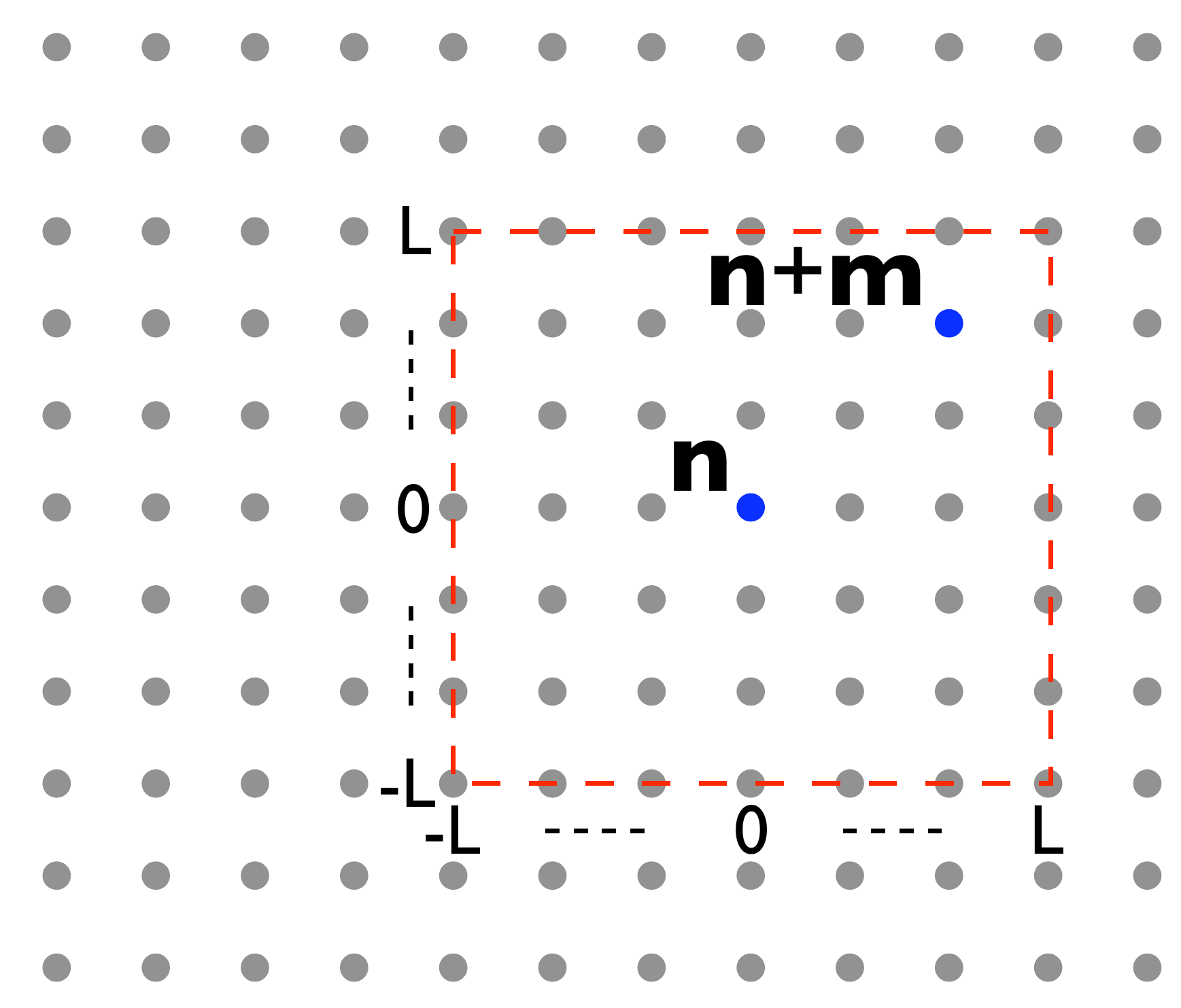}
 \caption{The non-zero hopping terms of $H_0$ link site ${\bf n}$ with sites that are confined in the square drawn with the dashed line. The square contains $(2L+1)\times (2L+1)$ number of sites.}
\end{figure}

We consider a tight-binding Hamiltonian $H_0$ defined on the 2-dimensional lattice ${\bf Z}^2$, which has $K$ number of quantum states per site. We use ${\bf n}$=$(n_1,n_2)$ to represent the sites of the lattice. The Hilbert space ${\mathcal H}$ is spanned by the vectors 
\begin{equation}
|{\bf n},\alpha\rangle, {\bf n} \in {\bf Z}^2, \alpha = 1,\ldots, K; \langle  {\bf n},\alpha| {\bf n}',\alpha' \rangle = \delta_{{\bf n},{\bf n'}}\delta_{\alpha \alpha'}.
\end{equation}
We adopt a very general form of the tight-binding Hamiltonian:
\begin{equation}
H_0=\sum_{{\bf n}\in {\bf Z}^2} \sum_{{\bf m}} \sum_{\alpha, \beta=1}^K [\Gamma_{\alpha \beta}^{\bf m} |{\bf n},\alpha\rangle \langle  {\bf n+m},\beta | + \Gamma_{\alpha \beta}^{{\bf m}*}|{\bf n+m},\beta\rangle \langle  {\bf n},\alpha |].
\end{equation}
The sum over ${\bf m}$ runs over a finite number of points of ${\bf Z}^2$. To be specific, we consider that these points are contained in a square of $(2L+1)\times (2L+1)$ number of sites, centered at the origin (see Fig.~1). An important assumption on $H_0$ is the existence of an insulating gap $\Delta$=$[E_-$,$E_+]$ in the spectrum. Under these conditions, the bulk system has the following general property.\medskip

\noindent {\bf Proposition 1.} The resolvent of $H_0$ has the exponential decay property:
\begin{equation}
\langle {\bf n},\alpha| (H_0-z)^{-1}|{\bf n}' ,\beta \rangle \leq \frac{e^{-q|{\bf n}-{\bf n}'|}}{\mbox{dist}(z,\sigma(H_0))-\zeta(q)},
\end{equation}
where $\zeta(q)$ is specified in Eq.~\ref{zeta}. $z$ is an arbitrary point in the complex energy plane, away from the spectrum $\sigma(H_0)$ of $H_0$. The above estimate holds only when $q$ is small enough so that the the denominator in the right hand side is positive.\medskip

We defer the technical proof to the Appendix. But we want to mention here that the exponential decay or localization of the resolvent away from the spectrum of the Hamiltonian is a very general property, which can be demonstrated through a variety of the techniques. For our concrete example, the property follows from a simple application of the Combes-Thomas principle \cite{Combes:1973nx}. The same principle can be used to obtain sharp exponential decay estimates for continuous Schrodinger operators in the presence of magnetic fields \cite{Prodan:2006cr}.   

\subsection{The system with a rough boundary}

\begin{figure}
 \center
 \includegraphics[width=7cm]{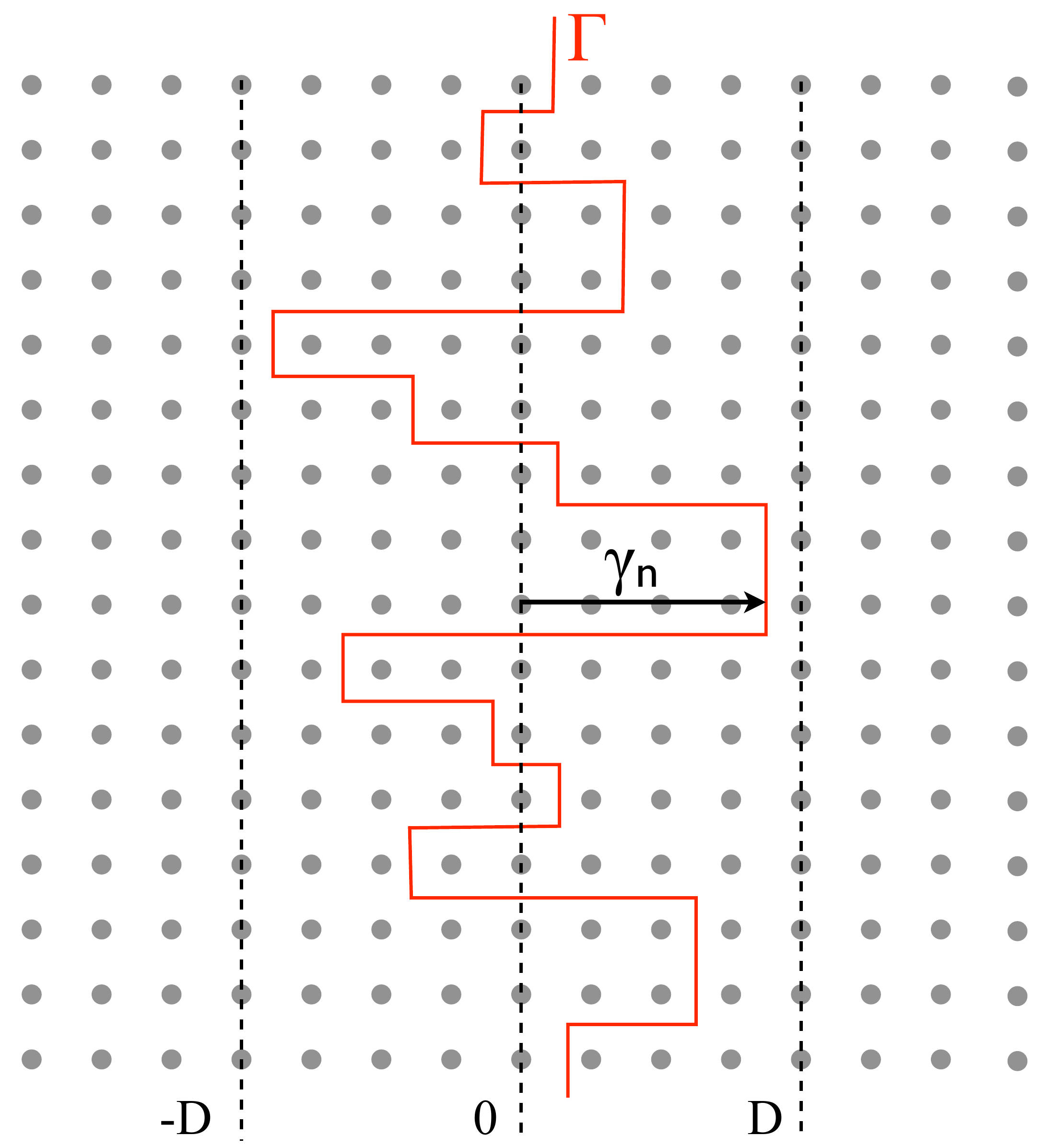}
 \caption{Exemplification of a rough edge $\Gamma$ and of the definition of $\gamma_n$.}
\end{figure}

Let us consider a line $\Gamma$ like in Fig.~2, where the only constraint we impose is that all points of $\Gamma$ be confined within a distance $D$ from the line $n_1$=0. The line $\Gamma$ can be described by a sequence $\{\gamma_n\}_n$, where $\gamma_n$ gives the deviation of $\Gamma$ from the axis $n_1$=0 at the row $n_2$=$n$ of our lattice, as exemplified in Fig.~2. We have $\gamma_n \in {\mathcal I}$, with ${\mathcal I}=\{-D+1/2,-D+3/2, \ldots,D-1/2\}$.  We recall that we use ${\bf n}$=$(n_1,n_2)$ to describe a point of the lattice. Thus, $\Gamma$ can be considered as a point of the set $\Omega={\mathcal I}^{\times \infty}$: $\Gamma$=$\{\ldots,\gamma_{-1},\gamma_0,\gamma_1,\ldots\}$. On the set $\Omega$, we introduce the product probability measure, denoted by $d\Gamma$, which is the infinite product of the simplest probability measure $\nu$ on ${\mathcal I}$: $\int f(n) d\nu(n)=\frac{1}{2D}\sum_{n\in {\mathcal I}} f(n)$, $f(n)$ being any function defined on ${\mathcal I}$. We remark that $d\Gamma$ obtained this way is ergodic relative to the discrete translations along the vertical direction of our lattice. We will use the probability measure $d\Gamma$ to average over all possible contours $\Gamma$.

We assign a sign to the points of the lattice, depending on their position relative to the curve $\Gamma$:
\begin{equation}
s_{\bf n}=\left \{
\begin{array}{l}
-1 \ \mbox{if {\bf n} is to the left of $\Gamma$} \\
+1 \ \mbox{if {\bf n} is to the right of $\Gamma$}
\end{array} \right .
\end{equation}
The Hilbert space ${\mathcal H}$ decomposes in a direct sum ${\mathcal H}={\mathcal H}_+ \oplus {\mathcal H}_-$, where
\begin{equation}
{\mathcal H}_\pm = span\{|{\bf n},\alpha\rangle, s_{\bf n} =\pm 1 \}
\end{equation}
We first introduce a very useful Hamiltonian $H$, which is obtained from $H_0$ by erasing the hopping terms that cross the contour $\Gamma$. Using the sign function introduced above, we can write the Hamiltonian $H$ as:
\begin{equation}
H=\sum_{s_{\bf n}\cdot s_{\bf n+m}>0} \sum_{\alpha \beta} [\Gamma_{\alpha \beta}^{\bf m} |{\bf n},\alpha\rangle \langle  {\bf n+m},\beta | + \Gamma_{\alpha \beta}^{{\bf m}*}|{\bf n+m},\beta\rangle \langle  {\bf n},\alpha |].
\end{equation}
It is also important to notice that $H = H_0 - \Delta V$ with:
\begin{equation}
\Delta V=\sum_{s_{\bf n}\cdot s_{\bf n+m}<0} \sum_{\alpha \beta} [\Gamma_{\alpha \beta}^{\bf m} |{\bf n},\alpha\rangle \langle  {\bf n+m},\beta | + \Gamma_{\alpha \beta}^{{\bf m}*}|{\bf n+m},\beta\rangle \langle  {\bf n},\alpha |].
\end{equation}
The sums involve only sites that are localized near the edge, more precisely those with the first coordinate $n_1$ in the interval [-$D$-$L$+1,$D$+$L$-1]. It is also interesting to notice that $H$ describes two decoupled systems with edges. Indeed, $H$ decomposes into a direct sum $H$=$H_-$$\oplus$$H_+$, where
\begin{equation}
\begin{array}{c}
  H_-:{\mathcal H}_-\rightarrow {\mathcal H}_- ,\medskip \\
H_-=\sum\limits_{s_{\bf n}, s_{\bf n+m}<0} \sum\limits_{\alpha \beta} [\Gamma_{\alpha \beta}^{\bf m} |{\bf n},\alpha\rangle \langle  {\bf n+m},\beta | + \Gamma_{\alpha \beta}^{{\bf m}*}|{\bf n+m},\beta\rangle \langle  {\bf n},\alpha |]
\end{array}
\end{equation}
and 
\begin{equation}
\begin{array}{c}
H_+:{\mathcal H}_+\rightarrow {\mathcal H}_+, \medskip \\
H_+=\sum\limits_{s_{\bf n} , s_{\bf n+m}>0} \sum\limits_{\alpha \beta} [\Gamma_{\alpha \beta}^{\bf m} |{\bf n},\alpha\rangle \langle  {\bf n+m},\beta | + \Gamma_{\alpha \beta}^{{\bf m}*}|{\bf n+m},\beta\rangle \langle  {\bf n},\alpha |].
\end{array}
\end{equation}
The system with the boundary $\Gamma$ is defined by the Hamiltonian $H_\Gamma$=$H_+$. We list three important properties of the system with the boundary.\medskip

\noindent {\bf Proposition 2.} Let $\phi(\epsilon)$ be a smooth function with support in the spectral gap of $H_0$. We consider that the support of $\phi$ is separated from the spectrum of $H_0$ by at least a distance $\delta$$>$0; $\delta$ will be considered fixed from now on.
\begin{enumerate}

\item Consider two lattice points, ${\bf n}$ and ${\bf n}'$ in the "+" side of the lattice. Then there exists $A(q)$$>$0, independent of ${\bf n}$ and ${\bf n}'$ and $\Gamma$, such that:
\begin{equation}\label{property1}
|\langle {\bf n},\alpha| \phi(H_\Gamma) | {\bf n}',\beta \rangle \leq A(q)e^{-q(n_1+n_1')},
\end{equation}
for $q$ small enough.

\item Fix a positive integer $N$. There exists $B_N$$>$0, independent of ${\bf n}$ and ${\bf n}'$ and $\Gamma$, such that:
\begin{equation}\label{property2}
|\langle {\bf n},\alpha| \phi(H_\Gamma) | {\bf n}',\beta \rangle \leq \frac{B_N}{(|{\bf n}-{\bf n}'|+1)^N}. \medskip
\end{equation}

\item There exists $C_N(q)>0$, independent of ${\bf n}$ and ${\bf n}'$ and $\Gamma$, such that:
\begin{equation}
|\langle {\bf n},\alpha| \phi(H_\Gamma) | {\bf n}',\beta \rangle| \leq C_N(q) \frac{e^{-q(n_1+n_1')}}{(|n_2- n_2'|+1)^N},
\end{equation}
where $q$ must be taken small enough. \medskip
\end{enumerate}

We defer the technical proof to the Appendix. Here we want to make the following comments. The first property states that $\phi(H_\Gamma)$ is exponentially localized near the boundary. This property follows from the simple identity:
\begin{equation}\label{identity00}
 R(z)=R_0(z)+R_0(z) \Delta V R_0(z)+R_0(z)\Delta V R(z) \Delta V R_0(z),
\end{equation}
where $R_0(z)$=$(H_0-z)^{-1}$ and $R(z)$=$(H-z)^{-1}$. The identity written in Eq.~\ref{identity00} reveals the following simple principle: if the resolvent of the bulk Hamiltonian has the exponential decay property, then the difference $R(z)-R_0(z)$ is exponentially localized near the edge, which is assured by the fact that $\Delta V$ is strictly localized near the boundary. Then the exponential localization of  $\phi(H_\Gamma)$ follows from Stone's formula:
\begin{equation}
\phi(H)  = \lim\limits_{\epsilon \rightarrow 0} \int\limits_{\bf R}  \mbox{Im} [R(\lambda+i\epsilon)-R_0(\lambda+i\epsilon)]\phi(\lambda)\frac{d \lambda}{\pi i},
\end{equation}
where the inclusion of $R_0$ has no effect, since the integral over $\lambda$ is constrained inside the spectral gap of $H_0$. We point out here that  $R(\lambda+i\epsilon)$ is evaluated infinitely closed to the spectrum which can be singular. The Appendix shows how to deal with this problem in a very general way.

The second property follows from an estimate on $H$ similar to that of Proposition 1 and from the rapid decay to zero of $\zeta(q)$ as $q$ goes to zero. The third property is a direct consequence of the previous two properties. As we shall see, these three properties assure that the minimal conditions stated in Ref.~\cite{Prodan:2008oq} are satisfied by the Chern insulators.

\section{The problem and the result.}

We define first the central observable,  the operator $\hat{y}_\Gamma$ defined on ${\mathcal H}_\Gamma$, which gives the vertical coordinate: 
\begin{equation}
\hat{y}_\Gamma|{\bf n},\alpha \rangle = n_2 |{\bf n},\alpha \rangle, \ {\bf n}=(n_1,n_2).
\end{equation}
The index $\Gamma$ is there to indicate that the operator is defined on ${\mathcal H}_\Gamma$. The observable $\hat{y}_\Gamma$ has discrete spectrum, $\sigma(\hat{y}_\Gamma)$=${\bf Z}$, each eigenvalue being infinitely degenerate. Let $\pi_\Gamma(n)$ denote the spectral projector of $\hat{y}_\Gamma$ onto the eigenvalue $n$. We have the following explicit expression:
\begin{equation}
\pi_\Gamma(n) = \sum\limits_{n_1 > \gamma_n, \alpha} |{\bf n},\alpha \rangle \langle {\bf n},\alpha |,
\end{equation}
where $\{\gamma_n\}_{n\in {\bf Z}}$ is the sequence corresponding to the contour $\Gamma$, as discussed in the previous Section.

On the large Hilbert space ${\mathcal H}$, we can implement the discrete lattice translations group along the vertical direction by:
\begin{equation}
u_n | (n_1,n_2),\alpha \rangle = | (n_1,n_2-n),\alpha \rangle.
\end{equation}
The discrete lattice translations along the vertical direction can also be extended to a map $t_n$ acting on the space $\Omega$ of all possible contours $\Gamma$. The map $t_n$ simply shifts a contour downwards by $n$ sites.

Let us collect now the important facts into the following list: \medskip

\begin{itemize}

\item We have defined the family of self-adjoint Hamiltonians $H_\Gamma$:${\mathcal H}_\Gamma$$\rightarrow$${\mathcal H}_\Gamma$,  $\Gamma \in \Omega$.

\item The observable $\hat{y}$ obeys:
\begin{equation}
u_n \hat{y}_\Gamma u_n^* = \hat{y}_{t_n \Gamma} + n
\end{equation} 

\item The family of Hamiltonians $H_\Gamma$ is covariant, namely, $u_n$ is an isometry that sends ${\mathcal H}_\Gamma$ into ${\mathcal H}_{t_n\Gamma}$ and $u_n H_\Gamma u_n^*=H_{t_n \Gamma}$.

\item On the set $\Omega$ we defined the probability measure $d\Gamma$, which is ergodic and invariant relative to the mappings $t_n$.

\end{itemize}

We now define the trace (notation $\mbox{tr}_0$) over the states of zero expectation value for $\hat{y}_\Gamma$:
\begin{equation}
\mbox{tr}_0 \{A\}=\mbox{Tr} \{\pi_\Gamma (0)A\pi_\Gamma(0)\},
\end{equation}
and we use $\mbox{tr}_0$ to define the current:
\begin{equation}
J_\Gamma =  \mbox{tr}_0 \left\{\rho(H_\Gamma) \frac{\mbox{d}\hat{y}_\Gamma(t)}{\mbox{d}t}\right\} =i \mbox{tr}_0 \left\{\rho(H_\Gamma) [H_\Gamma,\hat{y}_\Gamma ]\right\}.
\end{equation}
Here $\rho(\epsilon)$ is the statistical distribution of the quantum states. Since we are interested in the contribution to the current coming from the edge states, we assume that the support of $\rho(\epsilon)$ is entirely contained in the interval $[E_- +\delta, E_+ - \delta]$ and that $\int \rho(\epsilon)d\epsilon$=1. We assume that $\rho(\epsilon)$ is a smooth function. Note that $\mbox{tr}_0$ above is finite precisely because of the properties stated in Proposition 2.\medskip

\noindent {\bf Main Theorem.}  Let $F(\epsilon)\equiv \int_{\epsilon}^{\infty} \rho(\epsilon)$. Note that $F(\epsilon)$ is smooth and equal to 1 below $E_- +\delta$ and to 0  above $E_+ - \delta$; also $F'(\epsilon)$=$-\rho(\epsilon)$. We recall that $E_\pm$ are the edges of the spectral gap of the bulk Hamiltionian $H_0$. Using the spectral calculus, we define the following unitary operators:
\begin{equation}\label{uomega}
U_\Gamma = e^{-2 \pi i F(H_\Gamma)}.
\end{equation}
\medskip If $\pi_\Gamma^+$ is the projector onto the non-negative spectrum of ${\hat y}_\Gamma$, then:
\begin{equation}\label{main}
\int_\Omega d\Gamma \ J_\Gamma  = \frac{1}{2\pi}   \ \mbox{Index} \left \{\pi_\Gamma^+ U_\Gamma \pi_\Gamma^+ \right \}.
\end{equation} 
The Index is an integer number, independent of the shape of $\rho(\epsilon)$ or the contour $\Gamma$.\medskip

The proof of the statement is given in the following sections. We continue here with a discussion of the Index, more precisely its invariance property and how to compute it. The Index is defined on the class of Fredholm operators as:
\begin{equation}
\mbox{Ind} A=\dim  Ker[A] -\dim  Ker[A^*],
\end{equation}
for $A:X\rightarrow Y$, with $X$ and $Y$ two Hilbert spaces. The Index has several important properties:
\begin{enumerate}
\item $\mbox{Ind} A^* = - \mbox{Ind} A$.
\item $\mbox{Ind} BA = \mbox{Ind}B + \mbox {Ind} A$, for $A:X\rightarrow Y$ and $B:Y\rightarrow Z$ two Fredholm operators.
\item $\mbox{Ind} (A+C) = \mbox{Ind} A$ if $C$ is compact (in particular, if $C$ is finite rank).
\item The Index is invariant to norm-continuous deformations of the operators inside the Fredholm class.
\end{enumerate} 

\begin{figure}
 \center
 \includegraphics[width=7cm]{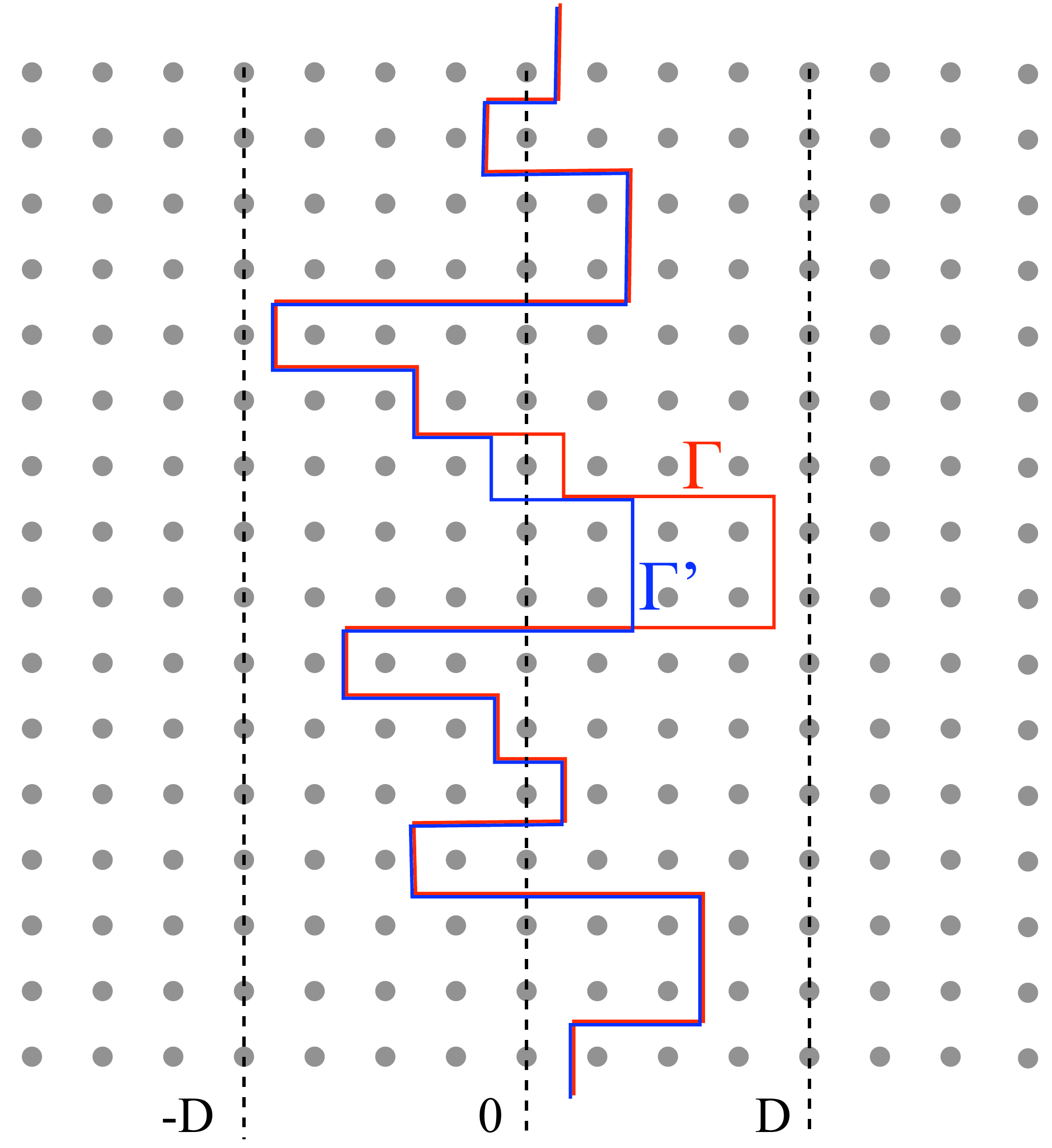}
 \caption{Example of a finite deformation of contours: $\Gamma \leftrightarrow \Gamma'$.}
\end{figure}

Based on these general properties, we argue that the Index is independent of the contour $\Gamma$. We consider a finite deformation as shown in Fig.~3, where the contour $\Gamma$ is deformed to the left, in a finite region, into the contour $\Gamma'$. We can equivalently say that $\Gamma'$ is deformed to the right into $
\Gamma$. In any case, ${\mathcal H}_{\Gamma} \subset {\mathcal H}_{\Gamma'}$. Let $i: {\mathcal H}_\Gamma \rightarrow {\mathcal H}_{\Gamma'}$ be the inclusion map and $p:{\mathcal H}_{\Gamma'} \rightarrow {\mathcal H}_{\Gamma}$ be the projection
\begin{equation}
p\sum\limits_{n_2\geq \gamma'_{n_2}} a_n |{\bf n},\alpha\rangle \langle {\bf n} \alpha | = \sum\limits_{n_2\geq \gamma_{n_2}} a_n |{\bf n},\alpha\rangle \langle {\bf n} \alpha |.
\end{equation}
Note that $p = i^*$, $p\circ i =1_{{\mathcal H}_\Gamma}$ and $i\circ p = \pi_{{\mathcal H}_\Gamma}$ (the projector from $H_{\Gamma'}$ to ${\mathcal H}_\Gamma$). Because the two curves differ only in a finite region, $\dim Ker [p]<\infty$ ($Ker[i]$ is empty) and the two operators are Fredholm. The insertion of $\pi_\Gamma^+ U_\Gamma \pi_\Gamma^+ $ in the ${\mathcal H}_{\Gamma'}$ given by $i\circ \pi_\Gamma^+ U_\Gamma \pi_\Gamma^+  \circ p$ has same Index since
\begin{equation}
\mbox{Ind} \ i \circ \pi_\Gamma^+ U_\Gamma \pi_\Gamma^+ \circ p = \mbox{Ind} \ i + \mbox{Ind} \ \pi_\Gamma^+ U_\Gamma \pi_\Gamma +   \mbox{Ind} \ i^*
\end{equation}
and the first Index and last Index cancel each other. The action of $i \circ \pi_\Gamma^+ U_\Gamma \pi_\Gamma^+ \circ p$ on the Hilbert space ${\mathcal H}_{\Gamma'}$ is very simple, 
\begin{equation}
\begin{array}{c}
i \circ \pi_\Gamma^+ U_\Gamma \pi_\Gamma^+ \circ p = (i \circ \pi_\Gamma^+ \circ p) ( i \circ U_\Gamma \circ p) (i \circ \pi_\Gamma^+ \circ p) \medskip \\
= (i \circ \pi_\Gamma^+ \circ p) e^{-2 \pi i F(i\circ H_\Gamma \circ p)} (i \circ \pi_\Gamma^+ \circ p).
\end{array}
\end{equation}
We notice first that the projectors $i \circ \pi_\Gamma^+ \circ p$ and $\pi_{\Gamma'}$ differ by a finite rank operator. Consequently,
\begin{equation}
\mbox{Ind} \ (i \circ \pi_\Gamma^+ \circ p) e^{-2 \pi i F(i\circ H_\Gamma \circ p)} (i \circ \pi_\Gamma^+ \circ p)
\end{equation}
and
\begin{equation} 
\mbox{Ind} \ \pi_{\Gamma'}^+ e^{-2 \pi i F(i\circ H_\Gamma \circ p)}  \pi_{\Gamma'}^+
\end{equation}
are the same. We also notice that $i\circ H_\Gamma \circ p$, acting on ${\mathcal H}_{\Gamma'}$, has the same expression as $H_\Gamma$, thus $i\circ H_\Gamma \circ p$ and $H_{\Gamma'}$ differ by a finite set of hopping terms. As the estimates of the next Sections will show, we can continuously switch on the missing hopping terms in $H_\Gamma$ until it becomes identical to $H_{\Gamma'}$ and keep
\begin{equation}
 \pi_{\Gamma'}^+  e^{-2 \pi i F(H_{\mbox{\tiny{deformed}}})} \pi_{\Gamma'}^+
 \end{equation}
 in the Fredholm class during the process. Due to the invariance of the Index under norm-continuous deformations, we can conclude
 \begin{equation}
 \mbox{Ind} \ \pi_{\Gamma} U_\Gamma  \pi_{\Gamma} =
 \mbox{Ind} \ \pi_{\Gamma'} U_{\Gamma'}  \pi_{\Gamma'}.
\end{equation}
Using successive deformations of the type describe above, we can always deform a contour into another and the conclusion is that the index is independent of $\Gamma$.

The Index can now be computed by taking the contour $\Gamma$ as a straight vertical line, in which case we can use the translational invariance, more precisely, the Bloch decomposition. Let us denote by $\Gamma_0$ such a straight contour. Definitely the theorem stated above applies equally well to the case when the set $\Omega$ reduces to one point, the contour $\Gamma_0$ (all we have to do is to take $D$=0). Then we have the following practical way of computing the Index:
\begin{equation}\label{index}
\begin{array}{c}
\mbox{Ind} \pi_{\Gamma_0}^+ U_{\Gamma_0} \pi_{\Gamma_0} = \mbox{Tr} \{ \pi_{\Gamma_0}(0) \rho(H_{\Gamma_0}) [H_{\Gamma_0},\hat{y}_{\Gamma_0}] \pi_{\Gamma_0}(0) \} \medskip \\
= \sum\limits_n \int\limits_{k=-\pi}^\pi  \rho(\epsilon_{n,k}) \partial_k \epsilon_{n,k} \ dk =  \sum\limits_n \alpha_n. 
\end{array}
\end{equation}
where $\epsilon_{n,k}$ are the edge bands and $\alpha_n$ are defined in Fig.~4 (we used here the fact that $\int \rho(\epsilon) d\epsilon$=1). In other words, the Index gives the difference between the number of forward and backward moving edge bands. This is a very simple but important result because the edge bands can be easily computed numerically for the homogenous case. Moreover, if necessary, one can further simplify the computation of the Index by considering continuous deformations of the bulk Hamiltonian that keep the insulating gap opened. Using the fundamental result of Kellendonk, Richter and Schulz-Baldes \cite{Kellendonk:2002p598}, we now know that the number computed in Eq.~\ref{index} is equal to the Chern number of the bulk states below the insulating gap.

\begin{figure}
 \center
 \includegraphics[width=9cm]{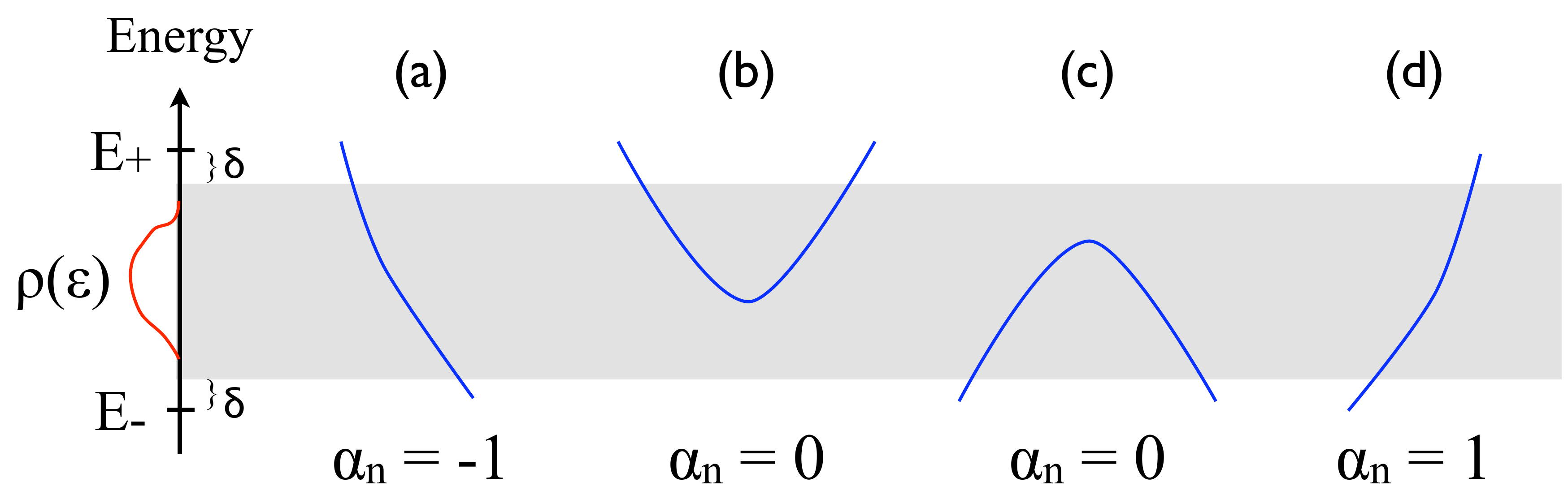}
 \caption{The figure lists all possible ways an edge band (blue lines) can cross the energy window $[E_- + \delta,E_+ - \delta]$. Below each configuration the figure presents the value of the coefficient $\alpha_n$.}
\end{figure}

As a final remark for this Section, note that our main statement is about the average of the edge current and not the current itself. However, since the family $\{H_\Gamma\}_{\Gamma \in \Omega}$ is covariant relative to translations, which act ergodically on $\Omega$,  the spectrum of $\{H_\Gamma\}$ is non-random. This implies that, if the edge spectrum becomes localized for a non-zero measure set of $\Omega$, it will be localized for all contours, except a possible zero measure set. But this cannot happen, exactly because the average of the edge current is non-zero for Chern insulators. This allows us to conclude that the rough edge cannot destroy the edge conducting channels.

\section{Proof of the Main Statement}

Before starting the proof of our main theorem, we collect a set of technical facts, which all follow from the estimates given in Propositions 1 and 2.

\subsection{Technical results}

 Along this paper, the following notations $\| \ \|$ and $\| \ \|_{\mbox{\tiny{HS}}}$ represent the operator and the Hilbert-Schmidt norms, respectively. \medskip
 
\noindent {\bf Proposition 3.} The following statements are true:
\begin{enumerate}

\item $(U_\Gamma-I) \pi_\Gamma (n)$ are Hilbert-Schmidt. Moreover, their Hilbert-Schmidt norm is less than an upper bound, which is independent of $\Gamma$.\\
\item With the notation
\begin{equation}
K_\Gamma (n,n')= \| \pi_\Gamma (n) (U_\Gamma-I)\pi_\Gamma (n') \| ^2 _{\mbox{\tiny{HS}}},
\end{equation}
there exists $G_N>0$, independent of $\Gamma$, sucht that
\begin{equation}
K_\Gamma (n,n') \leq \frac{G_N}{(1+|n-n'|)^N},
\end{equation}
for any positive integer $N$. \\

\item If $\pi_\Gamma^-$ denotes the projector onto the negative spectrum of $\hat{y}_\Gamma$ and $\Sigma_\Gamma \equiv \pi_\Gamma^+ - \pi_\Gamma^-$, then $[\Sigma_\Gamma, U_\Gamma]$ are Hilbert-Schmidt. Moreover, their Hilbert-Schmidt norm is less than an upper bound, which is independent of $\Gamma$.\\

\item $[\hat{y}_\Gamma,U_\Gamma]\pi_\Gamma(n)$ are Hilbert-Schmidt. Moreover, their Hilbert-Schmidt norm is less than an upper bound, which is independent of $\Gamma$.
\end{enumerate}\medskip 

Let us also state a fundamental property of tr$_0$, essential for the proof of our main statement.\medskip

\noindent{\bf Proposition 4.}  If $\{A_\Gamma\}_{\Gamma \in \Omega}$ and $\{B_\Gamma\}_{\Gamma\in \Omega}$ are two covariant families of operators such that $\pi_\Gamma(0) A_\Gamma$, $A_\Gamma \pi_\Gamma(0)$, $\pi_\Gamma(0) B_\Gamma$ and $B_\Gamma\pi_\Gamma(0)$ are Hilbert-Schmidt with uniformly bounded norms, then:
\begin{equation}
\int d\Gamma \ \mbox{tr}_0 \{A_\Gamma B_\Gamma \}=\int d\Gamma \ \mbox{tr}_0 \{B_\Gamma A_\Gamma \}<\infty.
\end{equation}\medskip

\medskip

We also need to introduce an approximate spectral projector onto the support of $\rho(\epsilon)$. For this, we consider a smooth function $G(\epsilon)$ which is equal to 0 below $E_-+\delta/2$, to 1/2 on the interval $[E_- +\delta,E_+ -\delta]$, and to 1 above $E_+-\delta/2$ (see Fig.~5). In this case,
\begin{equation}
\pi_s = \frac{1}{2}\left (I-e^{-2\pi i G(H_\Gamma)}\right)
\end{equation}
leaves invariant the states corresponding to the spectral interval of support of $\rho(\epsilon)$. By construction, $\pi_s$ has similar properties as $U_\Gamma - I$, which are listed stated below. Note that $\pi_s$ depends on $\Gamma$, in fact $\{\pi_s\}_\Gamma$ form a covariant family. The dependence on $\Gamma$ will not be stated explicitly. 

\begin{figure}
\center
 \includegraphics[width=9cm]{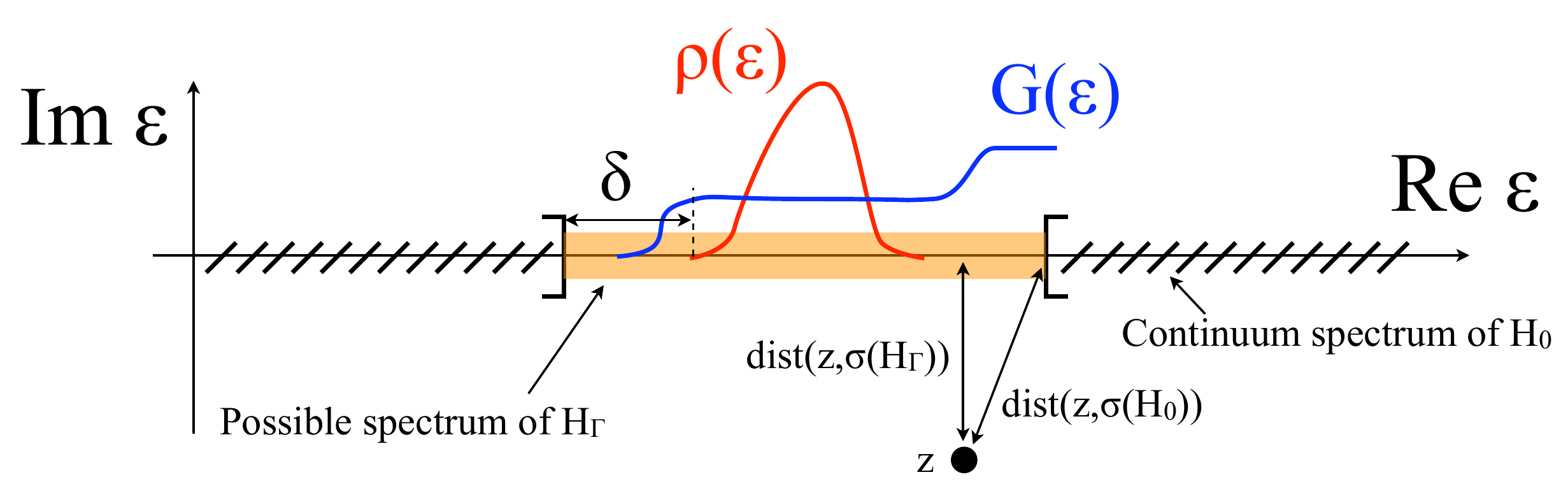}\\
 \caption{Illustration of several definitions used in the text.}  
 \end{figure}

\subsection{The Proof}

The proof makes use of the following non-commutative version of the Residue Theorem, whose proof can be found in Ref.~\cite{Prodan:2008oq}:\medskip

\noindent{\bf A Non-Commutative Residue Theorem.} Let $f(z)$ be analytic in a strip around the unit circle. If $\{U_\Gamma\}_{\Gamma \in \Omega}$ a covariant family of unitary operators such that $(U_\Gamma-I)\pi_\Gamma(0)$ and $[\hat{y}_\Gamma, U_\Gamma]\pi_\Gamma(0)$ are Hilbert-Schmidt, then:
\begin{equation}
\begin{array}{c}
\int d\Gamma \ \mbox{tr}_0\{(f(U_\Gamma)-f(I))[\hat{y}_\Gamma, U_\Gamma]\}
=b_1 \int d\Gamma \ \mbox{tr}_0 \{(U_\Gamma^*-I) [\hat{y}_\Gamma,U_\Gamma]\},
\end{array}
\end{equation}
where $b_1$ is the coefficient appearing in the Laurent expansion:
\begin{equation}
f(z)=\sum\limits_{n=1,\infty} b_n z^{-n} + \sum\limits_{n=0,\infty} a_n z^n.
\end{equation}\medskip

Now from Proposition 4 we know that $[\Sigma_\Gamma,U_\Gamma]$ is Hilbert-Schmidt. Then $\pi_\Gamma^+ U_\Gamma \pi_\Gamma^+$ is Fredholm and we can use Connes' result in non-commutative geometry:
\begin{equation}
\mbox{Ind}\{\pi_\Gamma^+ U_\Gamma \pi_\Gamma^+\}=-\frac{1}{4}\mbox{Tr}\{\Sigma_\Gamma[\Sigma_\Gamma,U_\Gamma^*][\Sigma_\Gamma,U_\Gamma]\},
\end{equation}
which, by using elementary manipulations \cite{Prodan:2008oq}, can be reformulated in a slightly different format:\medskip
\begin{equation}
\mbox{Ind}\{\pi_\Gamma^+ U_\Gamma \pi_\Gamma^+\}=-\frac{1}{2}\sum\limits_{\beta=\pm} \mbox{Tr}\{\pi_\Gamma^\beta (U_\Gamma^*-I)[\Sigma_\Gamma,U_\Gamma]\pi_\Gamma^\beta\}.   \medskip
\end{equation}
We use the projectors $\pi_\Gamma(n)$ to expand
\begin{equation}\label{expand}
\begin{array}{c}
\mbox{Ind}\{\pi_\Gamma^+ U_\Gamma \pi_\Gamma^+\}
=-\frac{1}{2}\sum\limits_{n} \mbox{Tr}\{\pi_\Gamma(n) (U_\Gamma^*-I)[\Sigma_\Gamma,U_\Gamma]\pi_\Gamma(n)\} .
\end{array}
\end{equation}
We now consider the average over $\Gamma$. Since the index is independent of $\Gamma$, the average over $\Gamma$ can be omitted for the left side. On the right hand side, we use the fact that the trace of trace-class operators is invariant to unitary transformations and the fact that the measure $d\Gamma$ is invariant to the mappings $t_n$, to write:
\begin{equation}\label{step1}
\begin{array}{c}
 \mbox{Ind}\{\pi_\Gamma^+ U_\Gamma \pi_\Gamma^+\}  \medskip \\
 =-\frac{1}{2}\int d\Gamma \sum\limits_{n}  \mbox{Tr}\{u_n\pi_\Gamma(n) (U_\Gamma^*-I)[\Sigma_\Gamma,U_\Gamma]\pi_\Gamma(n) u_n^* \} \medskip \\
=-\frac{1}{2}\int d\Gamma\sum\limits_{n} \mbox{Tr}\{\pi_{t_n\Gamma}(0) (U_{t_n\Gamma}^*-I)[u_n\Sigma_\Gamma u_n^*,U_{t_n\Gamma}]\pi_{t_n\Gamma}(0) \} \medskip \\
=-\frac{1}{2}\int d\Gamma\sum\limits_{n} \mbox{Tr}\{\pi_{\Gamma}(0) (U_{\Gamma}^*-I)[u_n\Sigma_{t_{-n}\Gamma} u_n^*,U_{\Gamma}]\pi_{\Gamma}(0) \}
\end{array}
\end{equation}
The key observation at this step is that:
\begin{equation}
\sum\limits_{n}u_{n}\Sigma_{t_{-n}\Gamma} u_{n}^*=2\hat{y}_\Gamma+I,
\end{equation}
whose graphical representation is given in Fig.~6, which leads to the intermediated conclusion that:
\begin{equation}\label{last}
\mbox{Ind}\{\pi _\Gamma^+U_\Gamma \pi_\Gamma^+\}=-\int d\Gamma \ \mbox{tr}_0\{ (U_{\Gamma}^*-I)[\hat{y}_\Gamma, U_{\Gamma}] \}.
\end{equation}
The integrand of the last integral is finite, fact that can be seen from Proposition 3. Before we go further we will do two things. First, we will place an approximate projector squared $\pi_s^2$ in front of $(U_{\Gamma}^*-I)$, which is legitimate since $\pi_s$ acts as identity on the space where $(U_{\Gamma}^*-I)$ is non-zero. It is easy to see that the conditions of the Proposition 4 are satisfied and for that reason we can move $\pi_s^2$ all the way to the right, inside $\mbox{tr}_0$. 

\begin{figure}
\center
 \includegraphics[width=8cm]{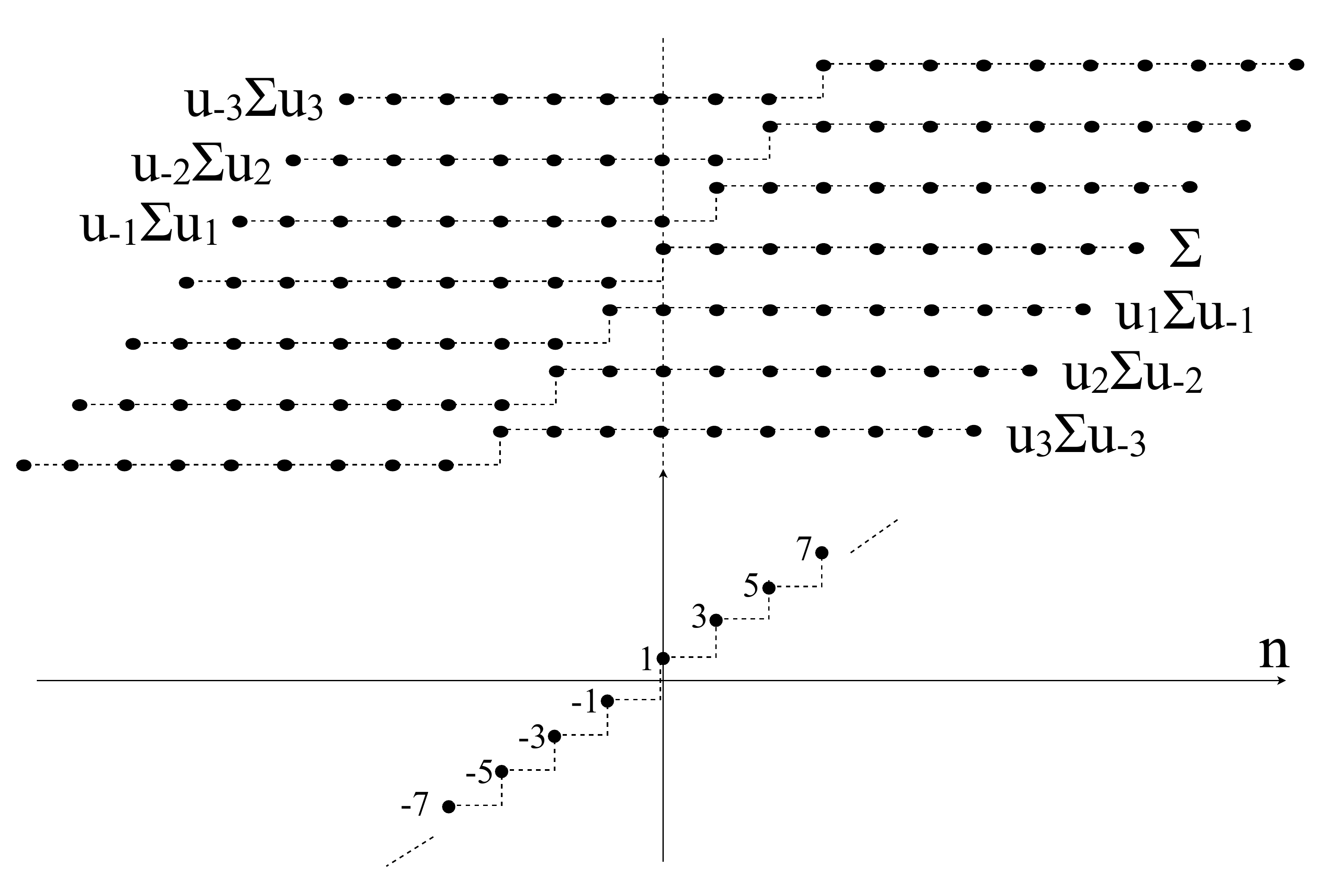}\\
 \caption{A graphical representation of $\sum_{n}u_{n}\Sigma_{t_{-n}\Gamma} u_{n}^*=\sum_{n}(2n+1)\pi_\Gamma(n)$. The top lines represent the spectral representations of $u_{x_n}\Sigma_{t_{-n}\Gamma} u_{x_n}^*$, which are discrete Heaviside functions shifted by $n$ sites. All these operators act on ${\mathcal H}_\Gamma$. The sum of the top lines results in the stair like function shown at the bottom.}  
 \label{fig2}
 \end{figure}

The second thing we do is,  by using the Non-Commutative Residue Theorem, replace $U^*_\Gamma - I$ by $\frac{1}{b_1}(f(U_\Gamma)-f(I))$, with $f(z)$ analytic in a strip around the unit circle. This step will become crucial at a later point in the proof, but it must be done at this step.
Thus, we arrived at
\begin{equation}\label{last}
\begin{array}{c}
 \mbox{Ind}\{\pi _\Gamma^+U_\Gamma \pi_\Gamma^+\}  =-\frac{1}{b_1} \int d\Gamma \ \mbox{tr}_0\{ (f(U_{\Gamma})-f(I))[\hat{y}_\Gamma, U_{\Gamma}]\pi_s^2 \}.
\end{array}
\end{equation}
 Like in Ref.~\cite{Kellendonk:2004p597}, we evaluate the commutator using Duhamel's identity
\begin{equation}
\begin{array}{c}
[\hat{y}_\Gamma, U_\Gamma(H_{\Gamma})]=-\int dt \tilde{\phi}(t) (1+it) \int_0^1 dq \times \medskip \\
e^{-(1-q)(1+it)H_\Gamma}[\hat{y}_\Gamma,H_\Gamma] e^{-q(1+it)H_\Gamma},
\end{array}
\end{equation}
with $\tilde{\phi}$ being the Laplace transform of $e^{-2\pi i F(x)}-1$, which is a smooth function with support in $[E_-+\delta,E_+-\delta]$. Then:
\begin{equation}
\begin{array}{c}
\mbox{Ind}\{\pi_\Gamma^+ U_\Gamma \pi_\Gamma^+ \} = \frac{1}{b_1}\int d\Gamma \int dt \tilde{\phi}(t) (1+it) \int_0^1 dq \times \medskip \\ 
\mbox{tr}_0\{(f(U_{\Gamma})-f(I))U_\Gamma 
e^{-(1-q)(1+it)H_\Gamma}[\hat{y}_\Gamma,H_\Gamma] \pi_s e^{-q(1+it)H_\Gamma} \pi_s\} .
\end{array}
\end{equation}
We will use now Proposition 4 to move $e^{-q(1+it)H_\Gamma}\pi_s$ all the way to the left, inside $\mbox{tr}_0$. Before doing that, we must be sure that the conditions of the Proposition 4 are satisfied. And indeed, the operators $e^{-q(1+it)H_\Gamma} \pi_s\pi_\Gamma(0)$ and $\pi_\Gamma(0) e^{-q(1+it)H_\Gamma} \pi_s$ are Hilbert-Schmidt, which follows from point (i) of Proposition 3 and the fact that $\pi_s$ has same properties as $U_\Gamma-I$. For the same reasons,
\begin{equation}
\pi_\Gamma(0) (f(U_{\Gamma})-f(I))U_\Gamma 
e^{-(1-q)(1+it)H_\Gamma}[\hat{y}_\Gamma,H_\Gamma] \pi_s
\end{equation}
and
\begin{equation}
(f(U_{\Gamma})-f(I))U_\Gamma 
e^{-(1-q)(1+it)H_\Gamma}[\hat{y}_\Gamma,H_\Gamma]\pi_s\pi_\Gamma(0)
\end{equation}
are Hilbert-Schmidt. After moving $e^{-q(1+it)H_\Gamma}\pi_s$ all the way to the left, inside $\mbox{tr}_0$,  we use the observation that all the operators to the left of $[\hat{y}_\Gamma,H_\Gamma]$ commute and that $\pi_s$ can be replaced with the identity, to obtain
 \begin{equation}
 \begin{array}{c}
\mbox{Ind}\{\pi_\Gamma^+ U_\Gamma \pi_\Gamma^+\} = \frac{1}{b_1}\int d\Gamma  \int dt \tilde{\phi}(t) (1+it) \int_0^1 dq \times \medskip \\ 
\mbox{tr}_0\{(f(U_{\Gamma})-f(I))U_\Gamma 
 e^{-(1+it)H_\Gamma}[\hat{y}_\Gamma,H_\Gamma] \pi_s \} \medskip \\
 = \frac{1}{b_1}\int d\Gamma \  
\mbox{tr}_0\{(f(U_{\Gamma})-f(I))\phi'(H_\Gamma)[\hat{y}_\Gamma,H_\Gamma] \pi_s \} \medskip \\
= \frac{ 2 \pi i}{b_1}\int d\Gamma \  
\mbox{tr}_0\{(f(U_{\Gamma})-f(I))U_\Gamma F'(H_\Gamma)[\hat{y}_\Gamma,H_\Gamma]  \pi_s\}  
\end{array}
\end{equation}
We now take
\begin{equation}
f(z)=\frac{z-1}{z-1+\epsilon}\frac{1}{z}, \ \epsilon>0,
\end{equation}
for which $b_1$=1. Then
\begin{equation}
\frac{1}{b_1}(f(U_\Gamma)-f(I))U_\Gamma=(U_\Gamma-I)(U_\Gamma-(1-\epsilon)I)^{-1},
\end{equation}
and by taking the limit $\epsilon \rightarrow 0$, we obtain:
\begin{equation}
\mbox{Ind}\{\pi_\Gamma^+ U_\Gamma \pi_\Gamma^+\} = 2 \pi i\int d\Gamma \   
\mbox{tr}_0\{\rho(H_\Gamma)[\hat{y}_\Gamma,H_\Gamma] \pi_s \}.
\end{equation}
The last equation is precisely the one of the theorem because $\pi_s$ acts as the identity over the space where $\rho(H_\Gamma)$ is non-zero.

\section{Conclusions}

We presented a fairly elementary proof that the edge channels of Chern insulators are not destroyed by a rough edge. In the process, we demonstrated the use of the new topological invariant introduced in Ref.~\cite{Kellendonk:2002p598} for solving interesting problems in condensed matter. 

The technical estimates derived in this paper hold for very general tight-binding Hamiltonians, in particular for lattice systems with spin. Thus the paper provides the technical ground for applications to spin-Hall effect. In other words, the formalism can be applied line by line to this case too, the only thing that remains to be identified is what observable should one use instead of $\hat{y}_\Gamma$ to get a non-zero value for the Index. This has been accomplished in Ref.~\cite{Prodan2}.

\section{Appendix}

\subsection{ Proof of Proposition 1.}

This will be an elementary application of the Combes-Thomas principle \cite{Combes:1973nx}. We do not seek here optimal estimates, which could be obtained with the technique developed in Ref.~\cite{Prodan:2006cr}, but rather the simplest proof of the above estimate. For this, consider the invertible transformation:
\begin{equation}
|{\bf n}\alpha\rangle \rightarrow U_{{\bf q}}|{\bf n},\alpha\rangle=e^{-{\bf q}{\bf n}}|{\bf n},\alpha \rangle, \ {\bf q}\in {\bf R}^2.
\end{equation}
With the notation $H_{{\bf q}}\equiv U_{-{\bf q}}H_0U_{{\bf q}}$, we have:
\begin{equation}
H_{{\bf q}}=\sum\limits_{{\bf n},{\bf m},\alpha,\beta} [\Gamma_{\alpha \beta}^{\bf m} e^{{\bf q}{\bf m}}|{\bf n},\alpha\rangle \langle  {\bf n+m},\beta |
+ \Gamma_{\alpha \beta}^{{\bf m}*}e^{-{\bf q}{\bf m}}|{\bf n+m},\beta\rangle \langle  {\bf n},\alpha |].
\end{equation}
We write $H_{\bf q}=H_0+W_{{\bf q}}$, with
\begin{equation}
\begin{array}{c}
W_{{\bf q}}=\sum\limits_{{\bf n},{\bf m},\alpha,\beta} [\Gamma_{\alpha \beta}^{\bf m} [e^{{\bf q}{\bf m}}-1]|{\bf n},\alpha\rangle \langle  {\bf n+m},\beta | \medskip \\
\ \ \ \ \ \ \ \ \ \ \ \ \ + \Gamma_{\alpha \beta}^{{\bf m}*}[e^{-{\bf q}{\bf m}}-1]|{\bf n+m},\beta\rangle \langle  {\bf n},\alpha |].
\end{array}
\end{equation}
Suppose we can show that $W_{{\bf q}}$ is small when $q$ is small. To be more precise, assume that the operator norm $\|W_{{\bf q}}\|$ goes to zero as $q$($\equiv |{\bf q}|$) goes to zero (recall that $\| A\| = \sup |\langle g | A| f \rangle |$, where supremum goes over all $\|f\|=\|g\|=1$). Then
\begin{equation}
\begin{array}{c}
\| H_{\bf q} -z\|=\|H_0-z + W_{\bf q}\| \medskip \\ 
\ge \|H_0-z\|-\|W_{\bf q}\|  \medskip \\
\ge \mbox{dist}(z,\sigma(H_0))-\|W_{\bf q}\|.
\end{array}
\end{equation}
The above inequality has substance only if the right hand side is positive, which is the case if $q$ is small enough. This estimate gives
\begin{equation}
\|(H_{\bf q}-z)^{-1}\leq \frac{1}{\mbox{dist}(z,\sigma(H_0))-\|W_{\bf q}\|},
\end{equation}
which becomes useful in the following way:
\begin{equation}
\begin{array}{c}
|\langle {\bf n},\alpha|(H_0-z)^{-1}|{\bf n}',\beta \rangle |e^{{\bf q}({\bf n}-{\bf n}'}) \medskip \\
=\langle {\bf n},\alpha|U_{\bf q} (H_0-z)^{-1}U_{-{\bf q}}|{\bf n}',\beta \rangle | \medskip \\
=|\langle {\bf n},\alpha| (H_{\bf q}-z)^{-1}|{\bf n}',\beta \rangle | \medskip \\
\leq (\mbox{dist}(z,\sigma(H_0))-\|W_{\bf q}\|)^{-1}.
\end{array}
\end{equation}
If we orient ${\bf q}$ parallel to ${\bf n}-{\bf n}'$ we obtain
\begin{equation}
|\langle {\bf n},\alpha|(H_0-z)^{-1}|{\bf n}',\beta \rangle | \leq 
\frac{e^{-q|{\bf n}-{\bf n}'|}}{\mbox{dist}(z,\sigma(H_0))-\sup_{|{\bf q}|=q}\|W_{\bf q}\|}.
\end{equation}
It remains to estimate $\|W_{\bf q}\|$. One can directly compute:
\begin{equation}
\langle {\bf n}+{\bf m},\beta|W_{{\bf q}}|{\bf n},\alpha \rangle = (e^{-{\bf qm}}-1)[\Gamma_{\beta \alpha}^{-{\bf m}}+\Gamma_{\alpha \beta}^{{\bf m}*}].
\end{equation}
We consider now two arbitrary unit vectors
\begin{equation}
\begin{array}{c}
f=\sum_{{\bf n},\alpha}a_{{\bf n},\alpha}|{\bf n},\alpha\rangle, \ g=\sum_{{\bf n},\alpha}b_{{\bf n},\alpha}|{\bf n},\alpha\rangle, \medskip \\ 
\sum_{{\bf n},\alpha}|a_{{\bf n},\alpha}|^2=\sum_{{\bf n},\alpha}|b_{{\bf n},\alpha}|^2=1,
\end{array}
\end{equation} 
and we use the Schwarz inequality to derive an upper bound for $\|W_{{\bf q}}\|$:
\begin{equation}
\begin{array}{c}
|\langle g |W_{{\bf q}}|f\rangle |=\left |\sum\limits_{{\bf n}',\beta} b^*_{{\bf n}',\beta} \langle {\bf n}',\beta|W_{{\bf q}}|f\rangle \right |
\leq \left [ \sum\limits_{{\bf n}',\beta} | \langle {\bf n}',\beta|W_{{\bf q}}|f\rangle|^2 \right ]^{1/2} \medskip \\
= \left [ \sum\limits_{{\bf n}',\beta} \left |\sum\limits_{{\bf n},\alpha} \langle {\bf n}',\beta|W_{{\bf q}}|{\bf n},\alpha\rangle a_{{\bf n},\alpha}\right |^2 \right ]^{1/2} \medskip \\
\leq \left [ K(2L+1)^2\sum\limits_{{\bf n}',\beta}  \sum\limits_{{\bf n},\alpha} |\langle {\bf n}',\beta|W_{{\bf q}}|{\bf n},\alpha\rangle|^2 |a_{{\bf n},\alpha} |^2 \right ]^{1/2}.
\end{array}
\end{equation}  
In the last step we used the fact that the number of non-zero terms in the sum over ${\bf n}$ and $\alpha$ cannot exceed $K(2L+1)^2$. We continue:
\begin{equation}
\begin{array}{c}
|\langle g |W_{{\bf q}}|f\rangle | \leq \left [ K(2L+1)^2  \sum\limits_{{\bf n},\alpha} \sum\limits_{{\bf m},\beta}\left |\langle {\bf n}+{\bf m},\beta|W_{{\bf q}}|{\bf n},\alpha\rangle|^2 |a_{{\bf n},\alpha}\right |^2 \right ]^{1/2} \medskip \\
= (2L+1)\sqrt{K} \left [\sum\limits_{{\bf n},\alpha} \sum\limits_{{\bf m},\beta}(e^{-{\bf qm}}-1)^2|\Gamma_{\beta \alpha}^{-{\bf m}}+\Gamma_{\alpha \beta}^{{\bf m}*}|^2 |a_{{\bf n},\alpha} |^2 \right ]^{1/2} \medskip \\
\leq (2L+1)\sqrt{K} \ \sup\limits_\alpha \left [\sum\limits_{{\bf m},\beta}(e^{-{\bf qm}}-1)^2|\Gamma_{\beta \alpha}^{-{\bf m}}+\Gamma_{\alpha \beta}^{{\bf m}*}|^2\right ]^{1/2}.
\end{array}
\end{equation}
We can then take
\begin{equation}\label{zeta}
\zeta(q) = (2L+1)\sqrt{K} \sup_{\alpha,|{\bf q}|=q}\left [\sum\limits_{{\bf m},\beta}(e^{-{\bf qm}}-1)^2|\Gamma_{\beta \alpha}^{-{\bf m}}+\Gamma_{\alpha \beta}^{{\bf m}*}|^2\right ]^{1/2},
\end{equation}
which evidently decays to zero as $q\rightarrow 0$.

\subsection{Proof of Proposition 2.}

(i) First, we point out that
\begin{equation}
|\langle {\bf n},\alpha| \phi(H_\Gamma) | {\bf n}',\beta \rangle=|\langle {\bf n},\alpha| \phi(H) | {\bf n}',\beta \rangle,
\end{equation}
if both ${\bf n}$ and ${\bf n}'$ are in the "+" zone of the lattice. Thus we can work with $H$ instead of $H_\Gamma$. Let us use the notation $R_0(z)$=$(H_0-z)^{-1}$ and $R(z)$=$(H-z)^{-1}$. We make use of the following simple identity:
\begin{equation}\label{identity1}
R(z)=R_0(z)+R_0(z) \Delta V R_0(z)+R_0(z)\Delta V R(z) \Delta V R_0(z),
\end{equation}
and we do the functional calculus via the Stone's formula:
\begin{equation}
\phi(H)  = \lim\limits_{\epsilon \rightarrow 0} \int\limits_{\bf R}  [R(\lambda+i\epsilon)-R(\lambda -i\epsilon)]\phi(\lambda)\frac{d \lambda}{2\pi i},
\end{equation}
where the limit is in the weak operator topology. Using Eq.~\ref{identity1} and the fact that the support of $\phi$ is in the gap of $H_0$, we obtain
\begin{equation}
\begin{array}{c}
\langle {\bf n},\alpha| \phi(H) | {\bf n}',\beta \rangle = \lim\limits_{\epsilon \rightarrow 0} \sum\limits_{{\bf m},{\bf k},\alpha',\beta'} \int\limits_{\bf R} \frac{d \lambda}{2\pi i} \ \phi(\lambda) \times \medskip \\
 \langle {\bf n},\alpha|R_0(\lambda)|{\bf m},\alpha' \rangle \langle {\bf k},\beta'| R_0(\lambda) | {\bf n}',\beta \rangle \times \medskip \\
 \langle{\bf m},\alpha'|\Delta V[ R(\lambda+i\epsilon)-R(\lambda -i\epsilon)]\Delta V | {\bf k},\beta' \rangle.
\end{array}
\end{equation}
It is important to notice that, due to the localization properties of $\Delta V$, the sum over ${\bf m}$ and ${\bf k}$ can be restricted to the sites with first coordinate within the interval [-$D$-$L$+1,$D$+$L$-1]. We consider the following vectors:
\begin{equation}
\psi_{{\bf m},\alpha'} = \Delta V |{\bf m},\alpha' \rangle, \ \  \psi_{{\bf k},\beta'} = \Delta V |{\bf k},\beta' \rangle.
\end{equation}
The norm of the two vectors is bounded by
\begin{equation}
\| \psi_{{\bf m},\alpha'}\|, \|\psi_{{\bf k},\beta'}\|\leq 2K(2L+1)^2\max \{ |\Gamma_{\alpha\beta}^{\bf m}| \}.
\end{equation}
We will call $Q$ the constant appearing to the right. With the notation
\begin{equation}
\Psi_{{\bf m},\alpha'}=\psi_{{\bf m},\alpha'}/\| \psi_{{\bf m},\alpha'}\|, \ \  \Psi_{{\bf k},\beta'}=\psi_{{\bf k},\beta'}/\|\psi_{{\bf k},\beta'}\|,
\end{equation}
we continue as follows:
\begin{equation}\label{inter1}
\begin{array}{c}
\langle {\bf n},\alpha| \phi(H) | {\bf n}',\beta \rangle \medskip \\
=\lim\limits_{\epsilon \rightarrow 0} \sum\limits_{{\bf m},{\bf k},\alpha',\beta'} \| \psi_{{\bf m},\alpha'}\| \|\psi_{{\bf k},\beta'}\| e^{-(q+\xi)(|{\bf n}-{\bf m}|+|{\bf n}'-{\bf k}|)} \int\limits_{\bf R} \frac{d \lambda}{2\pi i} \times \medskip \\
  \phi(\lambda)
 \langle {\bf n},\alpha|R_0(\lambda)|{\bf m},\alpha' \rangle e^{(q+\xi)|{\bf n}-{\bf m}|} \langle {\bf k},\beta'| R_0(\lambda) | {\bf n}',\beta \rangle e^{(q+\xi)|{\bf n}'-{\bf k}|}\times \medskip \\
 \langle\Psi_{{\bf m},\alpha'}|[ R(\lambda+i\epsilon)-R(\lambda -i\epsilon)]|\Psi_{{\bf k},\beta'} \rangle,
\end{array}
\end{equation}
where $q$$>$0 and $\xi$$>$0 are chosen so that Proposition 1 applies with $q$ replaced by  $q$+$\xi$. We consider the following function, 
\begin{equation}
\begin{array}{c}
F_{\bf nmkn'}^{\alpha\alpha'\beta'\beta}(\lambda)=\phi(\lambda)
 \langle {\bf n},\alpha|R_0(\lambda)|{\bf m},\alpha' \rangle e^{(q+\xi)|{\bf n}-{\bf m}|} \medskip \\
\times \langle {\bf k},\beta'| R_0(\lambda) | {\bf n}',\beta \rangle e^{(q+\xi)|{\bf n}'-{\bf k}|}
 \end{array}
\end{equation}
and observe that Eq.~\ref{inter1} can be written as:
\begin{equation}\label{inter2}
\begin{array}{c}
\langle {\bf n},\alpha| \phi(H) | {\bf n}',\beta \rangle 
= \sum\limits_{{\bf m},{\bf k},\alpha',\beta'} \| \psi_{{\bf m},\alpha'}\| \|\psi_{{\bf k},\beta'}\| \medskip \\ 
\times e^{-(q+\xi)(|{\bf n}-{\bf m}|+|{\bf n}'-{\bf k}|)} 
\int\limits_{\bf R}  F_{\bf nmkn'}^{\alpha\alpha'\beta'\beta}(\lambda) d\mu (\lambda)
\end{array}
\end{equation}
where $d\mu(\lambda)$ is the spectral measure of $H$, projected on $\Psi_{{\bf m},\alpha'}$ and $\Psi_{{\bf k},\beta'}$, which are unit vectors. Thus, we can obtain an upper bound in the following way:
\begin{equation}\label{inter3}
| \langle {\bf n},\alpha| \phi(H) | {\bf n}',\beta \rangle | 
\leq Q^2 \sum\limits_{{\bf m},{\bf k}}   
e^{-q(|{\bf n}-{\bf m}|+|{\bf n}'-{\bf k}|)} 
\sum\limits_{\alpha',\beta'} \sup\limits_\lambda  |F_{\bf nmkn'}^{\alpha\alpha'\beta'\beta}(\lambda) |.
\end{equation}
From Proposition 1 we have:
\begin{equation}
\sup\limits_\lambda  |F_{\bf nmkn'}^{\alpha\alpha'\beta'\beta}(\lambda) |\leq 
\frac{\sup_\lambda \phi(\lambda)}{[\delta - \zeta(q+\xi)]^2} 
\end{equation}
Also, for the remaining sums we have
\begin{equation}
\sum\limits_{{\bf m}}   e^{-(q+\xi)|{\bf n}-{\bf m}|} \leq S_\xi e^{-qn_1}, \ \  \sum\limits_{{\bf k}}   e^{-(q+\xi)|{\bf n}'-{\bf k}| }\leq S_\xi e^{-qn_1'}
\end{equation} 
where $S_\xi$ is a positive parameter depending only on $\xi$. If we put everything together we obtain:
\begin{equation}\label{prop1}
| \langle {\bf n},\alpha| \phi(H) | {\bf n}',\beta \rangle | 
\leq Q^2 K^2 S_\xi^2  \frac{\sup_\lambda \phi(\lambda)}{[\delta - \zeta(q+\xi)]^2}  e^{-q(n_1+n_1')}.
\end{equation} \medskip

\noindent (ii) Here we do the functional calculus using a technique introduced by Helffer and Sjostrand \cite{Helffer:1989kx}. We consider the domain ${\mathcal D}$ in the complex plane defined by all those $z$ with $|$Im$z|$$\leq$$v_0$ and Re$z \in [E_- +\delta,E_+ + \delta]$. We recall that the distance from the support of $\phi$ and the spectrum of $H_0$ is at least $\delta$. We will take $v_0$ less than this $\delta$. Since $\phi(\epsilon)$ is smooth, for any positive integer $N$, one can construct a function $f$:${\mathcal D}$$\rightarrow$${\bf C}$ such that 

a) $f_N(z,\bar{z})$=$\phi(z)$ when $z$ is on the real axis. 

b) $|\partial_{\bar{z}}f_N(z,\bar{z})\leq \alpha_N |\mbox{Im} z|^N$.\smallskip

\noindent Such function is called an almost analytic extension of $\phi$. Using such function, one has
\begin{equation}
\phi(H) = \frac{1}{2\pi}\int\limits_{\mathcal D} \partial_{\bar{z}}f_N(z,\bar{z}) (H-z)^{-1}d^2 z.
\end{equation}
It is easy to see that the result stated in Proposition 1 applies equally well to the resolvent of $H$:
\begin{equation}
\langle {\bf n},\alpha| (H-z)^{-1}|{\bf n}' \beta \rangle \leq \frac{e^{-q|{\bf n}-{\bf n}'|}}{\mbox{dist}(z,\sigma(H))-\zeta(q)},
\end{equation}
for $q$ small enough such that  the denominator is positive. The difference between the two cases is that $H$ may not have a spectral gap like $H_0$ (see Fig.~5). Still, the result is of interest to us since it gives the behavior of $R(\lambda)$ for $z$ in ${\mathcal D}$ and away from the real axis. One can repeat the estimates on $W_{\bf q}$, with the links crossing the contour $\Gamma$ erased, and convince himself that the function $\zeta(q)$ remains the same. If $z$=$u+iv$, then dist$(z,\sigma(H))$$\geq$$|v|$. Looking at the expression for $\zeta$ function given in Eq.~\ref{zeta}, we see that it behaves linearly with $q$ for small values of $q$. Thus, if we take $q$=$|v|/M$, with $M$ large enough, there is a $\theta >0$ such that 
 \begin{equation}
 \mbox{dist}(z,\sigma(H))-\zeta(|v|/M)>\theta |v|>0,
 \end{equation}
for any $z\in {\mathcal D}$. This gives us an upper bound on the resolvent for all $z$ in ${\mathcal D}$:
\begin{equation}
\langle {\bf n},\alpha| (H-u-iv)^{-1}|{\bf n}' \beta \rangle \leq \frac{e^{-|v||{\bf n}-{\bf n}'|/M}}{\theta |v|}.
\end{equation}
Then we can continue
\begin{equation}
\begin{array}{c}
| \langle {\bf n},\alpha| \phi(H) | {\bf n}',\beta \rangle | \medskip \\
=\left |\frac{1}{2\pi} \int du \int\limits_{-v_0}^{v_0}dv \ \partial_{\bar{z}}f_N(z,\bar{z}) \langle {\bf n},\alpha|(H-u-iv)^{-1} | {\bf n}',\beta \rangle \right | \medskip \\
\leq \frac{\alpha_N}{\pi \theta} \int du \int\limits_{0}^{v_0} dv \ v^{N-1} e^{-v|{\bf n}-{\bf n}'|/M} \medskip \\
\leq \frac{\alpha_N}{\pi \theta} e^{v_0/M} \int du \int\limits_{0}^{v_0} dv \ v^{N-1} e^{-v(|{\bf n}-{\bf n}'|+1)/M} \medskip \\
\leq \frac{\alpha_N \Delta M^N }{\pi \theta (|{\bf n}-{\bf n}'|+1)^N} e^{v_0/M} \int\limits_{0}^{v_0 (|{\bf n}-{\bf n}'|+1)/M} dx \ x^{N-1} e^{-x} \medskip \\
\leq \frac{\alpha_N \Delta M^N}{\pi \theta (|{\bf n}-{\bf n}'|+1)^N}e^{v_0/M} \int\limits_{0}^{\infty} dx \ x^{N-1} e^{-x} 
=  \frac{\alpha_N M^N \Gamma(N)}{\pi \theta (|{\bf n}-{\bf n}'|+1)^N} e^{v_0/M}.
\end{array}
\end{equation}

\noindent (iii) Replace $q$ by $2q$ at point (i) and and replace $N$ by $2N$ at point (ii) and take the product of Eqs.~\ref{property1} and \ref{property2} to obtain:
\begin{equation}
|\langle {\bf n},\alpha| \phi(H_+) | {\bf n}',\beta \rangle|^2 \leq A(2q)B_{2N} \frac{e^{-2q(n_1+n_1')}}{(|{\bf n}-{\bf n}'|+1)^{2N}}.
\end{equation} 
The statement follows because we can replace $|{\bf n}-{\bf n}'|$ by $|n_2-n_2'|$ any time.

\subsection{Proof of Proposition 3.}

\noindent {\it Proof.} (i) We will show that the Hilbert-Schmidt norms are uniformly bounded:
\begin{equation} 
\| (U_\Gamma-I) \pi_\Gamma (n)\|_{\mbox{\tiny{HS}}}^2 
=\mbox{Tr} \{\pi_\Gamma (n) (U_\Gamma^*-I) (U_\Gamma-I) \pi_\Gamma (n) \}.
\end{equation}
Since 
\begin{equation}
(U_\Gamma^*-I) (U_\Gamma-I)= (e^{2 \pi i F(H_\Gamma)}-1)(e^{-2 \pi i F(H_\Gamma)}-1)
\end{equation}
and the function $(e^{2 \pi i F(x)}-1)(e^{-2 \pi i F(x)}-1)$ is smooth and with support in the interval $[E_- +\delta,E_+ - \delta]$, we can apply Proposition 2 to conclude at this step that
\begin{equation}
\langle {\bf n},\alpha | (U_\Gamma^*-I) (U_\Gamma-I) | {\bf n}',\beta\rangle \leq A(q) e^{-q(n_1+n_1')}.
\end{equation}
Then
\begin{equation}
\begin{array}{c}
\mbox{Tr} \{\pi_\Gamma (n) (U_\Gamma^*-I) (U_\Gamma-I) \pi_\Gamma (n) \} \medskip \\
= \sum\limits_{n_1>\gamma_n,\alpha} \langle {\bf n},\alpha | (U_\Gamma^*-I) (U_\Gamma-I) |{\bf n},\alpha \rangle  \medskip \\
\leq K \sum\limits_{n_1>\gamma_n} A(q)e^{-2q n_1} \medskip \\
\leq K \sum\limits_{n_1>-D} A(q)e^{-2q n_1}\leq \infty.
\end{array}
\end{equation}

\noindent (ii) 
\begin{equation}
\begin{array}{c}
\| \pi_\Gamma (n) (U_\Gamma-I) \pi_\Gamma (n')\|_{\mbox{\tiny{HS}}}^2 \medskip \\
=\mbox{Tr} \{\pi_\Gamma (n') (U_\Gamma^*-I) \pi_\Gamma(n)(U_\Gamma-I) \pi_\Gamma (n') \} \medskip \\
=\sum\limits_{n_1>\gamma_n} \sum\limits_{n_1'>\gamma_{n'}}\sum\limits_{\alpha,\beta}
|\langle (n_1,n),\alpha | (U_\Gamma-I)|(n_1',n'),\beta\rangle|^2
\end{array}
\end{equation}
and by applying Proposition 2 point (iii) we have
\begin{equation}
\begin{array}{c}
\| \pi_\Gamma (n) (U_\Gamma-I) \pi_\Gamma (n')\|_{\mbox{\tiny{HS}}}^2 \medskip \\
\leq K^2\sum\limits_{n_1>\gamma_n} \sum\limits_{n_1'>\gamma_{n'}}
A_N(q)^2 \frac{e^{-2q(n_1+n_1')}}{(|n-n'|+1)^{2N}} \medskip \\
\leq\frac{K^2}{(|n-n'|+1)^{2N}} \left ( \sum\limits_{n_1>-D} 
A_N(q) e^{-q n_1}\right)^2.
\end{array}
\end{equation}

\noindent (iii) We proceed as follows.
\begin{equation}
\begin{array}{c}
\| [\Sigma_\Gamma,U_\Gamma] \|_{\mbox{\tiny{HS}}}^2 = \sum\limits_{n,n'} \| \pi_\Gamma(n) [\Sigma_\Gamma,U_\Gamma]\pi_\Gamma(n') \|_{\mbox{\tiny{HS}}}^2 \medskip \\
=\sum\limits_{n\cdot n' \leq 0} \| \pi_\Gamma(n) (U_\omega-I)\pi_\Gamma(n') \|_{\mbox{\tiny{HS}}}^2 = \sum\limits_{n\cdot n' \leq 0} K_\Gamma (n,n').
\end{array}
\end{equation}
In the last two sums, we must exclude the term $n$=$n'$=0. If we take $N>3$ at point (ii), the final sum is convergent and uniformly bounded.\smallskip

\noindent (iv) We use the following equivalent expression for the commutator $[\hat{y}_\Gamma,U_\Gamma]$:
\begin{equation}
\begin{array}{c}
[\hat{y}_\Gamma,U_\Gamma]=\sum\limits_{n,n'}  (n-n') \pi_\Gamma(n) (U_\Gamma-I) \pi_\Gamma(n')
\end{array}
\end{equation}
to obtain:
\begin{equation}
\begin{array}{c}
\| [\hat{y}_\Gamma,U_\Gamma]\pi_\Gamma(n) \|_{\mbox{\tiny{HS}}}^2  = \sum\limits_{n'}  |n-n'|^2  \| \pi_\Gamma(n) (U_\Gamma-I) \pi_\Gamma(n') \|_{\mbox{\tiny{HS}}}^2 \ \ \nonumber \medskip \\
 =  \sum\limits_{n'}  |n-n'|^2 K_\Gamma(n,n').
\end{array}
\end{equation} 
If we take $N>3$ at point (ii), the final sum is convergent and uniformly bounded.

\subsection{Proof of Proposition 4.}

\noindent{\it Proof.} Let $\pi (M)=\sum\limits_{-M}^M \pi_\Gamma(n)$. Then $\pi_\Gamma(0) A_\Gamma  \pi(M) B_\Gamma \pi_\Gamma(0)$ are trace class and
\begin{equation}
\begin{array}{c}
\int d\Gamma \ \mbox{Tr}\{\pi_\Gamma(0) A_\Gamma  \pi(M) B_\Gamma \pi_\Gamma(0) \} \medskip \\
=\int d\Gamma \ \sum\limits_{n=-M}^M\mbox{Tr}\{\pi_\Gamma(0) A_\Gamma  \pi_\Gamma(n) B_\Gamma \pi_\Gamma(0) \} \medskip \\
=\int d\Gamma \ \sum\limits_{n=-M}^M\mbox{Tr}\{\pi_\Gamma(n) B_\Gamma  \pi_\Gamma(0) A_\Gamma \pi_\Gamma(n) \} 
\end{array}
\end{equation}
At this point we use the invariance of the trace (on trace class operators) under the unitary transformations to continue:
\begin{equation}
\begin{array}{c}
\ldots =\int d\Gamma \ \sum\limits_{n=-M}^M \mbox{Tr}\{u_n\pi_\Gamma(n) B_\Gamma  \pi_\Gamma(0) A_\Gamma \pi_\Gamma(n) u_n^*\} \medskip \\
=\int d\Gamma \sum\limits_{n=-M}^M \mbox{Tr}\{\pi_{t_n\Gamma}(0) B_{t_n\Gamma}  \pi_{t_n\Gamma}(n) A_{t_n\Gamma} \pi_{t_n\Gamma}(0)\} \medskip \\
=\int d\Gamma \sum\limits_{n=-M}^M \mbox{Tr} \{\pi_\Gamma(0) B_\Gamma \pi_\Gamma(n) A_\Gamma  \pi_\Gamma(0)\}.
\end{array}
\end{equation}
At the end of above argument we used the invariance of $d\Gamma$ relative to the transformations $t_n$. We then have that
\begin{equation}
\begin{array}{c}
\int d\Gamma \ \mbox{Tr}\{\pi_\Gamma(0) A_\Gamma  \pi(M) B_\Gamma \pi_\Gamma(0) \} \medskip \\
=\int d\Gamma \  \mbox{Tr}\{\pi_\Gamma(0) B_\Gamma  \pi(M) A_\Gamma \pi_\Gamma(0) \},
\end{array}
\end{equation}
and the affirmation follows by letting $M$ go to infinity.

\bibliographystyle{my-h-elsevier}

\medskip


\begin{thebibliography}{10}
\expandafter\ifx\csname url\endcsname\relax
  \def\url#1{{\tt #1}}\fi
\expandafter\ifx\csname urlprefix\endcsname\relax\def\urlprefix{URL }\fi
\providecommand{\eprint}[2][]{\url{#2}}

\bibitem{Hatsugai:1993cs}
Hatsugai Y 1993 {\em Phys. Rev. B\/} {\bf 48} 11851--11862

\bibitem{Hatsugai:1993jt}
Hatsugai Y 1993 {\em Phys. Rev. Lett.\/} {\bf 71} 3697--3700

\bibitem{Kellendonk:2002p598}
Kellendonk J, Richter T and Schulz-Baldes H 2002 {\em Rev. Math. Phys\/} {\bf
  14} 87--119

\bibitem{SchulzBaldes:2000p599}
Schulz-Baldes H, Kellendonk J and Richter T 2000 {\em J. Phys. A: Math. Gen\/}
  {\bf 33} L27--L32

\bibitem{Kellendonk:2004p597}
Kellendonk J and Schulz-Baldes H 2004 {\em J. of Func. Analysis\/} {\bf 209}
  388--413

\bibitem{Elbau:2002qf}
Elbau P and Graf G 2002 {\em Comm. Math. Phys.\/} {\bf 229} 415--432

\bibitem{Combes:2005qd}
Combes J and Germinet F 2005 {\em Comm. Math. Phys.\/} {\bf 256} 159--180

\bibitem{Elgart:2005rc}
Elgart A, Graf G and Schenker J 2005 {\em Comm. Math. Phys.\/} {\bf 259}
  185--221

\bibitem{Haldane:1988dz}
Haldane F~D~M 1988 {\em Phys. Rev. Lett.\/} {\bf 61} 2015--2018

\bibitem{Sheng:2006vn}
Sheng D, Weng Z~Y, Sheng L and Haldane F 2006 {\em Phys. Rev. Lett.\/} {\bf 97}
  036808

\bibitem{Thonhauser:2006kx}
Thonhauser T and Vanderbilt D 2006 {\em Phys. Rev. B\/} {\bf 74} 23511

\bibitem{Haldane:2008ys}
Haldane F and Raghu S 2008 {\em Phys. Rev. Lett.\/} {\bf 100} 013904

\bibitem{Prodan:2008oq}
Prodan E 2009 {\em J. Phys. A: Math and Th.\/} {\bf 42} 065207

\bibitem{Prodan2}
Prodan E 2009 {\em J. Phys. A: Math and Th.\/} in press.

\bibitem{Combes:1973nx}
Combes J and Thomas L 1973 {\em Comm. Math. Phys.\/} {\bf 34} 251--270

\bibitem{Prodan:2006cr}
Prodan E, Garcia S and Putinar M 2006 {\em J. Phys. A: Math. Gen\/} {\bf 39}
  389--400

\bibitem{Helffer:1989kx}
Helffer B and Sjostrand J 1989 {\em Lecture Notes in Physics\/} vol 345
  (Springer) pp 118--197

\end{thebibliography}

\end{document}